\documentclass[aps,amsmath,amssymb,prl,superscriptaddress,twocolumn,showpacs,floatfix]{revtex4}
\usepackage{graphicx}
\usepackage[english]{babel}

\def\agetcn{$\kappa$-(BE\-DT\--TTF)$_2$\-Ag$_2$(CN)$_3$}
\def\etcn{$\kappa$-(BE\-DT\--TTF)$_2$\-Cu$_2$(CN)$_3$}

\def\kcucn{$\kappa$-CuCN}
\def\kagcn{$\kappa$-AgCN}
\def\EMdmit{$\beta'$-EtMe$_3$Sb[Pd(dmit)$_2$]$_2$}
\def\etme3{EtMe$_3$Sb}
\def\dmit{[Pd(dmit)$_2$]$_2$}
\def\Betaetme3{$\beta'$-EtMe$_3$Sb}
\def\betadmit{$\beta'$-[Pd(dmit)$_2$]$_2$}

\def\etal{{\it et al.}}

\begin{document}
\title{Importance of van der Waals interactions and cation-anion coupling in an organic quantum spin liquid}
\author{P. Lazi\'{c}}
\affiliation{Rudjer Bo\v{s}kovi\'{c} Institute, Bijeni\v{c}ka cesta 54, HR-10000 Zagreb, Croatia}
\author{M.\ Pinteri\'{c}}
\affiliation{Institut za fiziku, P.O.Box 304, HR-10001 Zagreb, Croatia}
\affiliation{Faculty of Civil Engineering, Smetanova 17, SI-2000 Maribor, Slovenia}
\author{D.\ Rivas G\'{o}ngora}
\affiliation{Institut za fiziku, P.O.Box 304, HR-10001 Zagreb, Croatia}
\author{A.\ Pustogow}
\affiliation{1.\ Physikalisches Institut, Universit\"{a}t Stuttgart, 
D-70550 Stuttgart, Germany}
\author{K.\ Treptow}
\affiliation{1.\ Physikalisches Institut, Universit\"{a}t Stuttgart, 
D-70550 Stuttgart, Germany}
\author{T.\ Ivek}
\affiliation{Institut za fiziku, P.O.Box 304, HR-10001 Zagreb, Croatia}
\author{O.\ Milat}
\affiliation{Institut za fiziku, P.O.Box 304, HR-10001 Zagreb, Croatia}
\author{B.\ Gumhalter}
\affiliation{Institut za fiziku, P.O.Box 304, HR-10001 Zagreb, Croatia}
\author{N. Do\v{s}li\'{c}}
\affiliation{Rudjer Bo\v{s}kovi\'{c} Institute, Bijeni\v{c}ka cesta 54, HR-10000 Zagreb, Croatia}
\author{M.\ Dressel}
\affiliation{1.\ Physikalisches Institut, Universit\"{a}t Stuttgart, 
D-70550 Stuttgart, Germany}
\author{S.\ Tomi\'{c}}
\affiliation{Institut za fiziku, P.O.Box 304, HR-10001 Zagreb, Croatia}
\email{stomic@ifs.hr}
\homepage{http://sceinlom.ifs.hr/}

\date{\today}

\begin{abstract}
The Mott insulator $\beta'$-EtMe$_3$Sb[Pd(dmit)$_2$]$_2$ belongs to a class of charge transfer solids with highly-frustrated triangular lattice of $S=1/2$ molecular dimers and a quantum-spin-liquid ground state. Our experimental and {\it ab initio} theoretical studies show the fingerprints of strong correlations and disorder,
important role of cation-dimer bonding in charge redistribution, no sign of intra- and inter-dimer dipoles,
and the decisive van der Waals contribution to inter-dimer interactions and the ground state structure. The latter consists of quasi-degenerate electronic states related to the different configurations of cation moieties which permit two different equally probable orientations. Upon reducing the temperature, the low-energy excitations slow down, indicating glassy signatures as the cation motion freezes out.
\end{abstract}


\maketitle

\label{sec:Intro}

Quantum spin liquid (QSL), a highly correlated fluctuating quantum spin state, is a long-standing intriguing phenomenon in physics \cite{Anderson1973,BalentsNature2010}. It is expected to appear in geometrically frustrated systems when strong quantum fluctuations suppress the long-range magnetic order. Recently the QSL has been realized in materials with frustrated lattices at the insulating side of Mott transition \cite{SavaryBalentsRPP2017ZhouKanodaRMP2017}. Because of this exciting discovery considerable efforts
have been devoted to the studies of organic Mott insulators - charge transfer crystalline solids in which electrons are strongly correlated and confined to two dimensions.
They form layers with triangular structures of molecular pairs--dimers with an odd number of electrons separated by non-conducting inorganic moities.

Three organic Mott systems with different degrees of correlations \cite{Pustogow2017} \etcn{} (short \kcucn{}), \agetcn{} (\kagcn{}) and \EMdmit{} (\Betaetme3{}) exhibit an anomalous electrodynamic response below 60\,K \cite{Pustogow2017,Abdel-Jawad,ItohPRL2013,Pinteric2014DresselRC2016,Pinteric2016}, while at very low temperatures the magnetic and thermodynamic response exhibits unconventional behavior ascribed to the QSL ground state \cite{ShimizuPRL2003Isono2016,ShimizuPRL2016,ItouWatanabe}. The full understanding of the nature of QSL is missing primarily because frustrated triangular lattices
on their own are unable to destroy the long-range magnetic order \cite{Huse1988}.
It was suggested that an additional, exotic spin-dipolar coupling in the presence of geometrical frustration would suffice to induce quantum fluctuations and suppress magnetic ordering, however its experimental confirmation is still lacking
\cite{Hotta,Ishihara,Sumit,Gomi,Sedlmeier2012}.
The mechanism of dipolar-spin coupling relies on the Coulomb interactions within the sublattice of molecular dimers only, thereby completely neglecting the role of inorganic moieties to which the molecular sublattice is strongly hydrogen-bonded. For \kcucn{} and \kagcn{} the decisive role of cation-anion hydrogen bonding is revealed by the experimental results combined with {\it ab initio} numerical calculations. This points to a significant anion-driven renormalization of the electronic properties due to disorder by cyanide isomorphism \cite{Pinteric2014DresselRC2016,Pinteric2016}. Notably, despite the fact that single crystals of these two BEDT-TTF materials are nominally pure, the variable-range hopping, relaxor dielectric response and anomalous terahertz response are observed within the molecular planes as a result of the entangled charge and lattice excitations.

\begin{figure}
\centering
\includegraphics[clip=true,width=0.78\columnwidth]{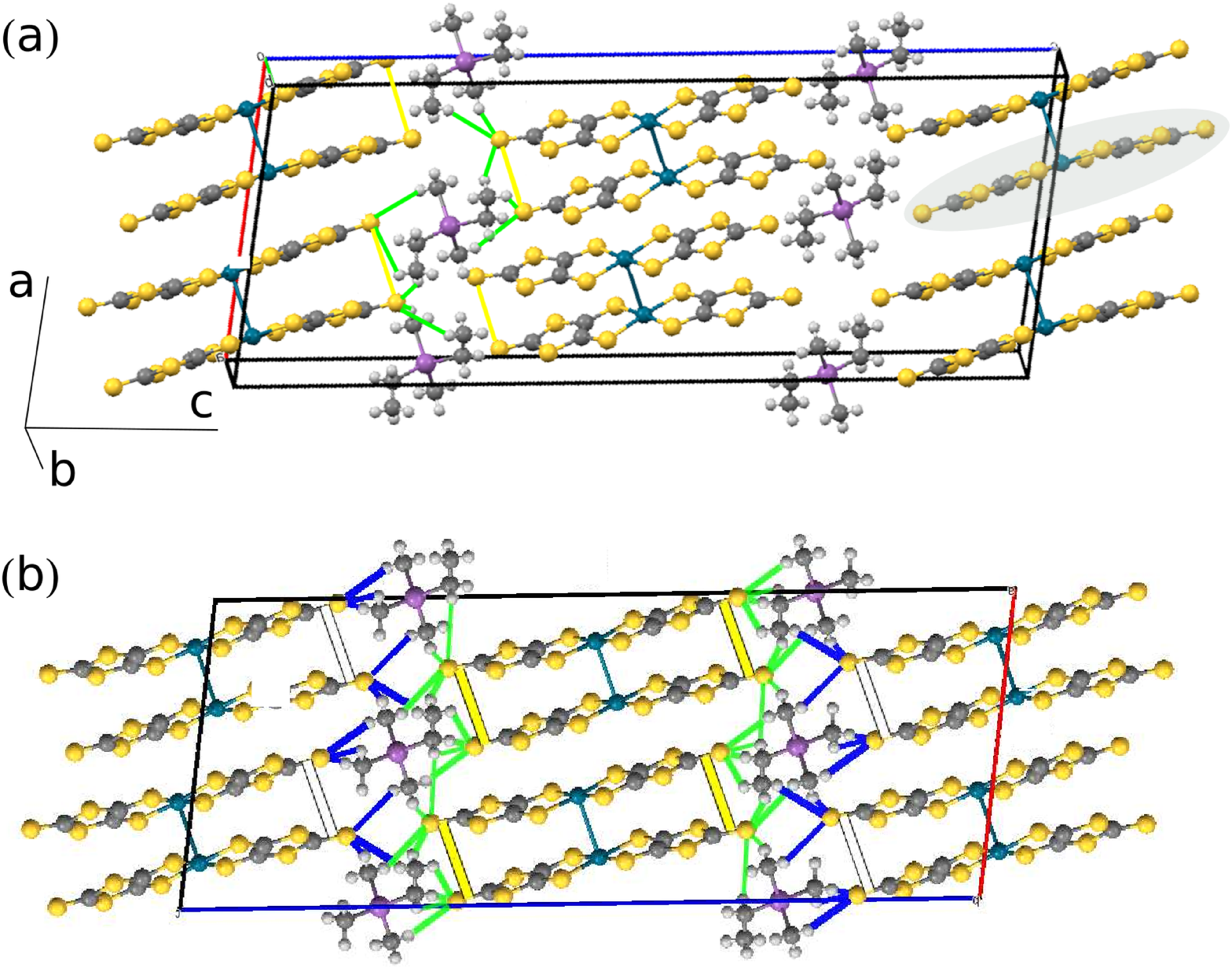}
\caption{(Color online) (a) Side view of extended unit cell in $C2/c$ symmetry of \Betaetme3{}. The molecule Pd(dmit)$_2$ is shown within  the shaded area. 
Yellow, grey, light grey and green circles denote sulfur, carbon, hydrogen and palladium atoms, respectively. H-bonds between the terminal S atoms of Pd(dmit)$_2$ molecules and H of Et (CH$_2$-CH$_3$) or Me (CH$_3$) groups of cations are marked by green lines. (b) Relaxed ground state structure in the $ac$ plane projected along the $b$-axis as obtained from DFT calculations. Et and the Me sites are in the proximity of S atoms of dimers in the central and side layer, respectively.
There are 8 short contacts (green lines) between two S atoms at one end of dimer and cations in the central layer, while there are only 6 contacts (blue lines) in the side layer. Yellow and white lines denote large and small S-S dimer openings near the Et and Me groups, respectively.}
\label{fig:structurecomposite}
\end{figure}

Important question thus arises on the role of random disorder in QSL formation. Does it only modify the QSL and makes the interpretation of experiments more difficult, or is it inherently related to the establishment of QSL \cite{FurukawaPRL2015}?
In order to shed more light on the mechanism of QSL entanglement in organic Mott insulators, and having in mind the proposed role of dipolar-spin coupling, we have investigated the QSL organic material \Betaetme3{} that consists of different molecular pairs and non-conducting ionic moieties
(Fig.\ \ref{fig:structurecomposite}(a), Fig.\ S1) \cite{KatoPSS2012}. Evidence for QSL is provided by the susceptibility and nuclear magnetic resonance measurements, which show no sign of classical phase transition down to 30\,mK but indicate the presence of an effective antiferromagnetic exchange interaction of the order of 250\,K \cite{ItouWatanabe}.
In this work we advance the understanding of this QSL Mott insulator by our results of dc transport, low-frequency dielectric and vibrational spectroscopy, and supporting density functional calculations (DFT). This allows us to unambiguously identify the origin of anomalous charge response in the cation disorder and motion mapped by hydrogen bonding onto the molecular dimers. We find that the ground state consists of quasi-degenerate electronic states with reduced symmetry the formation of which is most appropriately considered as a cooperative effect of the \etme3{} cation - \dmit{} anion coupling.
This reveals that the effective inter-dimer interaction
results from the interplay between attractive van der Waals interactions and Pauli repulsion.

\begin{figure}
\includegraphics[clip,width=0.75\columnwidth]{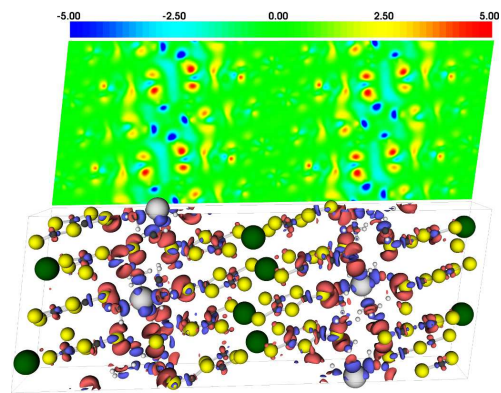}
\caption{(Color online) Lower panel: Cation-anion interaction-induced electronic charge redistribution in the relaxed ground state calculated as the difference between the charge density of the complete system and its constituents, i.e., cations and anions. Two isosurfaces are shown: electron accumulation (red) and depletion (blue). Balls of different color denote different atoms in the structure. The largest electron accumulation is concentrated close to the end S sites of \dmit{} dimers, it is much more enhanced near Et than Me sites. Upper panel: the averaged charge density in the $ac$ plane \cite{SM}. The color bar shows charge transfer ($e/$\AA{}$^2$).}
\label{fig:charge-distribution}
\end{figure}

Randomness in \Betaetme3{} is revealed
by x-ray diffraction measurements \cite{Kato2012}. These data show that Et groups of four cations in the unit cell occupy one of the two different equally probable orientations [Fig.\ \ref{fig:structurecomposite}(a)], indicating that the formation of several crystallographic configurations is possible (see Figs.\ S2--S4) \cite{SM}. To get an insight as how the randomness affects the electronic properties of \Betaetme3{} we have performed DFT calculations by using the self-consistently implemented nonlocal van der Waals density functional (vdW-DF) \cite{Dion04,Roman09,Klimes11} for correlation and optB88 for exchange \cite{Mittendorfer}. In that sense, our approach differs from the DFT calculations applied previously \cite{Nakamura2012,ScrivenPRL2012,Tsumuraya2013,Jacko2013}.  The entire \Betaetme3{} system, with atomic positions and unit cell parameters, relaxes into a structure of minimum energy when both ends of molecular dimers in each of the two layers in the unit cell are in the same cation environment. The space group of the relaxed ground state structure is reduced to P$\bar{1}$.
In the center and side layer the dimer ends are next to the Et and Me groups, respectively [Fig.\ \ref{fig:structurecomposite}(b), Fig.\ \ref{fig:charge-distribution}, lower panel].
Here it is worth noting that DFT relaxation without vdW interaction yields a structure with a 22\% larger unit cell than found experimentally, whereas the vdW-DF relaxed structure
meets the experimental one to within 4\%.
The inadequacy of DFT calculations without the vdW interactions is especially striking in the dimer plane: the $a$ and $b$ unit cell parameters are enlarged by almost 10\% (Table SI), thus indicating the significant role of nonlocal vdW binding contribution to the effective inter-dimer interactions within the $ab$ plane, as well as to the total cohesion energy of \Betaetme3{} crystal. This failure of the semi-local DFT was not so apparent in the \kcucn{} and \kagcn{} systems where
the structure was dominantly determined by the networks of bound anions Cu$_2$(CN)$_3^-$ and Ag$_2$(CN)$_3^-$ \cite{Pinteric2014DresselRC2016,Pinteric2016}. Furthermore we analyze the bonding of the system through charge rearrangement between its components.
We find the largest electron accumulation in \dmit{} dimers concentrated at the terminal S sites where chemical binding dominates. Also, it is much more enhanced near Et sites than near Me sites (Fig.\ \ref{fig:charge-distribution}). Accordingly, dimers establish stronger bonding with cations in Et environment than they do near Me groups. While
all the contacts are shorter than the sum of van der Waals radii, the bonds are larger in number and shorter in length for Et than for Me case [Fig.\ \ref{fig:structurecomposite}(b)].

\begin{figure}[h]
\includegraphics[clip,width=0.995\columnwidth]{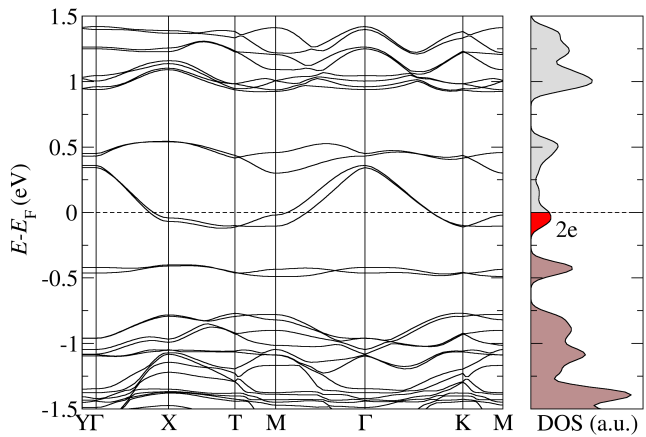}
\caption{(Color online) Calculated band structure in the relaxed ground state of \Betaetme3{} plotted along the high-symmetry directions. The segments $\Gamma X$, $\Gamma K$ and $\Gamma Y$ in the first Brillouin zone correspond to the $a$-, $b$- and $c$-direction, respectively. Weakly dispersive bands are identified at about 0.5\,eV, -0.4\,eV and -1.1\,eV, while strongly dispersive bands cross the Fermi level and accomodate 2 electrons 
(red shaded area in the density of states).}
\label{fig:band_structure}
\end{figure}

The electronic band structure 
for the relaxed state with minimum energy (i.e., the ground state) is shown in Fig.\ \ref{fig:band_structure}. Globally, and apart from fine details near the Fermi energy ($E_\mathrm{F}$), it resembles the bands obtained in previous calculations using the standard density functionals and fixed crystallographic parameters \cite{Nakamura2012,ScrivenPRL2012,Tsumuraya2013,Jacko2013}. While due to omission of appropriate electronic correlations all the band calculations find a half-filled band straddling $E_\mathrm{F}$, the angle-resolved photoemission measurements identify the lower Hubbard band near $E_\mathrm{F}$, and a soft Mott gap \cite{Ge2014}. In addition to strongly dispersive bands crossing $E_\mathrm{F}$, we point out two non-dispersive bands which are situated at the energies around -0.4 and -1\,eV and corresponding closely to the bands observed in ARPES along the $\Gamma X$ segment. Moreover, the overall band structure around $E_\mathrm{F}$ is dominantly anion-derived. This is suggested by the result for the band structure of an isolated self-standing anion subsystem which is only rigidly upshifted in energy if compared to the band structure of the whole anion-cation system (Fig.\ S9). 
Finally, we find that the band structure remains almost unchanged under pressure, confirming the pertinent role of vdW forces (
Fig.\ S10).

In addition to the relaxed state with minimum energy, we identify a manifold of electronic states in the energy range between 28\,meV and 146\,meV above the ground state. Remarkably, the two lowest energy states are associated with the configurations of $P\bar{1}$ and $P{1}${} symmetry, respectively, in which the two sides of dimers in each layer are in the same cation environment (either Et or Me), whereas their environment alternates in the central and side layers. The band structures of these quasi-degenerate configurations are very similar. Conversely, the states with highest energy are associated with configurations of $C_c$ symmetry in which the two sides of dimers are in alternate cation environment
, while the environment in two layers is the same. Between these two extremes we find the states with energies ranging from 45 to 80\,meV with the symmetry $P_{21}/n$. 
Such a strong dependence on the environment can be recognized as a result of competition between attractive vdW and Pauli repulsive contributions to the interdimer interaction within the $ab$ planes. The ground state configuration reflects the dominance of vdW interaction over Pauli repulsion whereas the reverse holds true for less probable configurations of higher energy. Importantly, the quasi-degenerate electronic ground state identified by the calculations implies a random domain structure in \Betaetme3{} that is qualitatively similar to \kcucn{} and \kagcn{} \cite{Pinteric2014DresselRC2016,Pinteric2016}.

\begin{figure}
\includegraphics[clip,width=0.6\columnwidth]{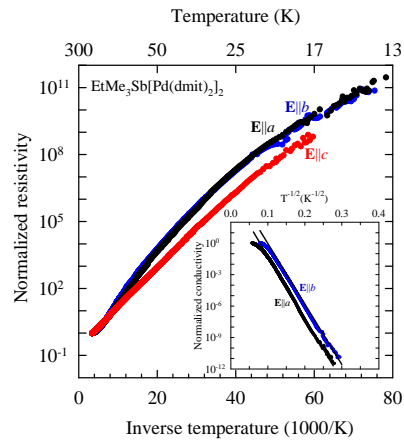}
\caption{(Color online) DC resistivity of \Betaetme3{} versus inverse temperature indicating nearest-neighbor hopping at higher temperatures. Inset displays normalized dc conductivity within dimer planes as a function of $T^{-1/2}$, demonstrating the Efros-Shklovskii hopping due to strong electron correlations for $T \lesssim 100$\,K. Data for $\mathbf{E}\parallel b$ are upshifted for clarity.}
\label{fig:dc}
\end{figure}

\begin{figure}
\includegraphics[clip,width=0.9\columnwidth]{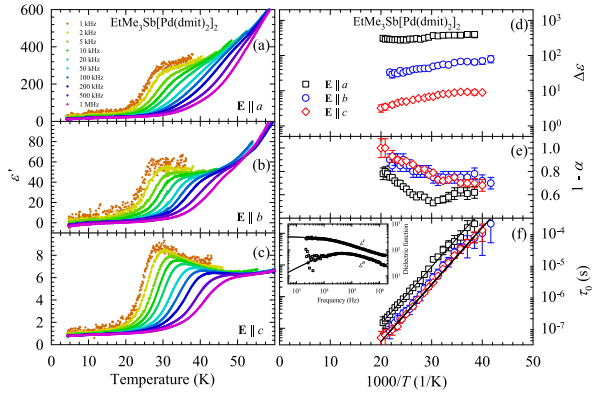}
\caption{(Color online) (a), (b), (c): Real part of the dielectric function $\varepsilon'$
of \Betaetme3{} versus temperature demonstrates the relaxor-like ferroelectric behavior for ac electric field applied within ($\mathbf{E}\parallel a$, $\mathbf{E}\parallel b$) and perpendicular ($\mathbf{E}\parallel c$) to the dimers plane. Dielectric strength $\Delta\varepsilon$ (d), relaxation time distribution ($1-\alpha$) (e), mean relaxation time $\tau_0$ (f) as function of inverse temperature. Inset in (e) shows double logarithmic plot of frequency dependence of real $\varepsilon'$ and imaginary $\varepsilon''$ parts of dielectric function for $\mathbf{E}\parallel a$ at 36\,K. Full lines are fits to generalized Debye function $\varepsilon(\omega)-\varepsilon_{\mathrm{HF}} = {\Delta\varepsilon}/[{1+({\rm i} \omega \tau_0)^{1-\alpha}}]$, $\varepsilon_{\mathrm{HF}}$ is high-frequency dielectric constant. Full lines in (f) show the Arrhenius slowing down of $\tau_0$ at the rate of 36\,meV along all three crystallographic axes.}
\label{fig:dielectric}
\end{figure}

In the following we measure the electronic properties and compare them with results of {\it ab initio} calculations. To this end we explore long-wavelength charge excitations by dielectric and dc-transport measurements performed along the three crystallographic axes. We first examine the dc transport and find fingerprints of randomness and strong correlations in accord with the results of DFT calculations (Fig.\ \ref{fig:dc}).
Especially, within the planes and at temperatures below $\sim\,100$\,K the hopping follows the Efros-Shklovskii temperature dependence $\sigma(T) \propto \exp\left[-(T_0/T)^{1/2}\right]$. In accord with ARPES \cite{Ge2014} this result indicates
a soft gap due to strong Coulomb interaction in disordered \Betaetme3{} \cite{EfrosShklovskii1975}. Here the electronic correlations are significantly stronger than in \kcucn{} and \kagcn{} as indicated by the optical data \cite{Pustogow2017}. The effective Coulomb interaction is comparable to intra-dimer coupling \cite{McKenzie1998}, accordingly the gap in the calculated generic anion bands (Fig.\ S9) exceeds the values for \kcucn{} and \kagcn{} \cite{Pinteric2014DresselRC2016,Pinteric2016}.”

Next we focus on the dielectric function extracted from the complex conductance measured as a function of temperature (300\,K -- 4.2\,K) and frequency (40\,Hz -- 10\,MHz) \cite{Pinteric2014DresselRC2016,Pinteric2016,SM}. On cooling, the real part of the dielectric function $\varepsilon'$ exhibits the relaxor-like ferroelectric behavior along the $a$ and $b$ axes within the dimer planes, as well as along the perpendicular $c$-axis (Fig.\ \ref{fig:dielectric}) \cite{Abdel-Jawad}. $\varepsilon'$ decays down to $\sim 50$\,K where the frequency dispersion starts to develop a peak which for the lowest frequencies is situated at $\sim 25$\,K. The approach based on simultaneous fits of $\varepsilon'$ and imaginary part $\varepsilon''$ of $\varepsilon(\omega)$ to the generalized Debye form reveals additional fingerprints of relaxors: a broad spectrum whose width $1-\alpha$ increases upon cooling and a gradual Arrhenius-like slowing down of the relaxation time $\tau_0$ (Fig.\ \ref{fig:dielectric}). Interestingly, the activation energy is similar along all the three crystallographic axes and comparable with dc resistivity, indicating the dominant free carrier screening.
Extrapolation to 100\,s gives the glass transition around 14\,K.

\begin{figure}
\includegraphics[clip,width=\columnwidth]{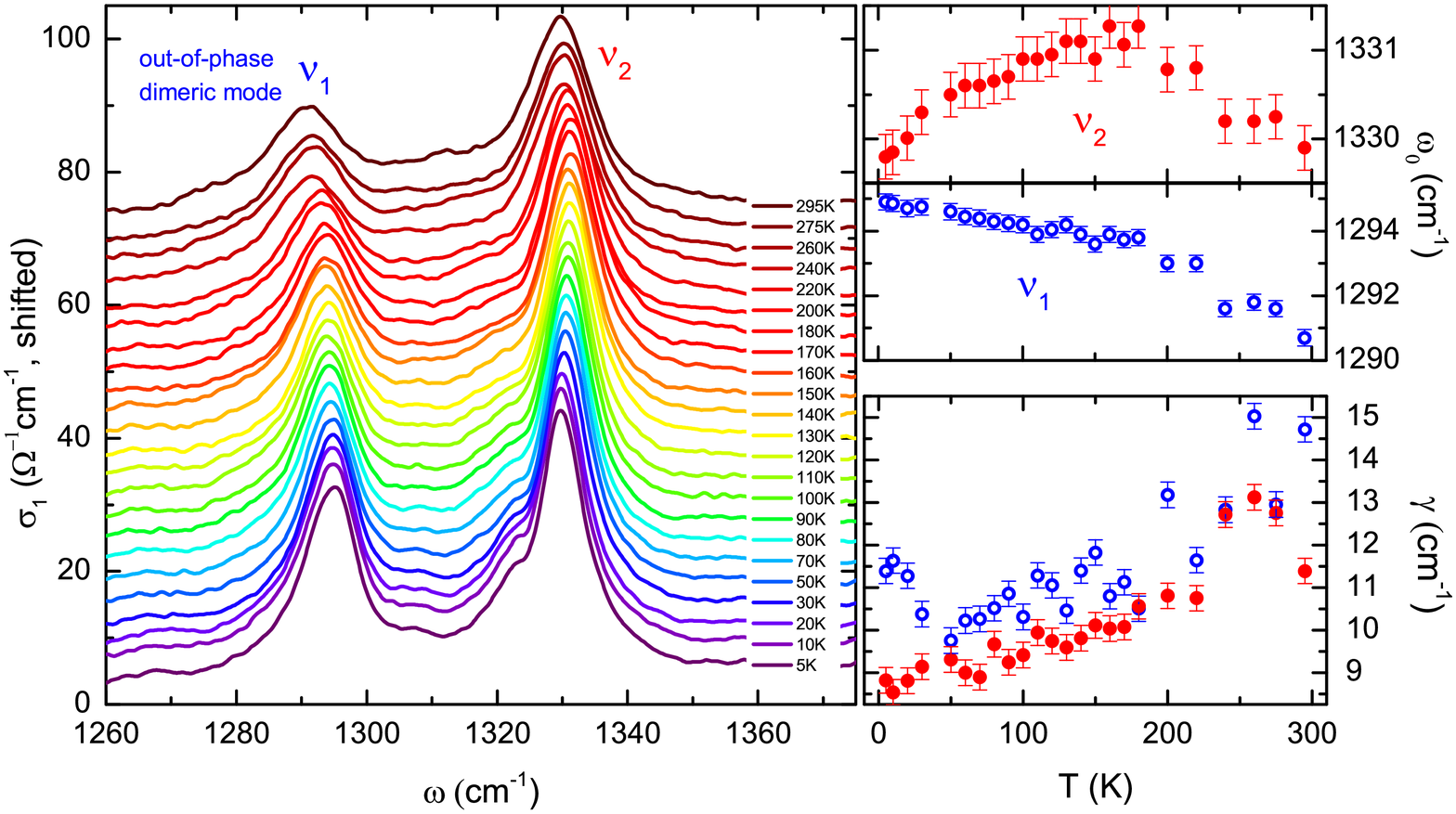}
\caption{(Color online)(a) Temperature evolution of the dimeric $\nu_1$ and intramolecular vibration $\nu_2$ of \Betaetme3{} measured for $\mathbf{E}\parallel c$. Except the 5\,K data, all curves are upshifted vertically for clarity. No splitting of the mode occurs. Temperature dependence of the resonance frequency (b) and damping (c) obtained by Fano function fits of the modes.}
\label{fig:nu2and1-caxis}
\end{figure}

The relaxor response suggests the existence of domains of low symmetry in accord with the DFT calculations. Basically, it allows two mechanisms of dielectric response ascribed either to the thermally activated reorientation of dipole moments within the domains, or to the motion of interphase boundaries. The latter mechanism was recently proposed to explain anomalous dielectric response in strongly correlated molecular solids \cite{Fukuyama2017}.
To test if the dipoles emerge upon cooling in \Betaetme3{} we employed vibrational spectroscopy to probe the most charge sensitive intramolecular $\nu_2$ and dimeric $\nu_1$ vibrations \cite{Yamamoto2011,SM}. The data in Fig.\ \ref{fig:nu2and1-caxis}(a) show no appreciable change besides the common thermal dependence. Most importantly, no splitting of either mode occurs down to 5\,K, ruling out the formation of molecular and dimeric dipoles and allowing only for charge fluctuations which are reduced when the temperature is lowered, likewise in \kcucn{} \cite{Sedlmeier2012}. However, the surprising $\nu_2$ temperature-dependent frequency shift draws attention. On cooling the mode shows first the commonly observed hardening,
whereas below 160\,K it displays a redshift getting steeper below 50\,K, right in the temperature range where the dielectric peak emerges. At these temperatures cation motion including the reorientation of Et groups and internal rotations of the Me groups
is present, as suggested by H/D-NMR study \cite{Kato_NMR}. DFT calculations of the potential energy profile for Me rotation give the energy barrier of 60\,meV and the first excited vibrational state only 11\,meV above the ground state \cite{SM}. Thus, as temperature lowers this motion is gradually hindered and due to the anion-cation coupling the freezing of cation motion is mapped by hydrogen bonding onto the conducting anions. Indeed, the relaxor response is also observed in \betadmit{} systems with uniquely defined Me positions \cite{Abdel-Jawad}.
The above scenario correlates well with elastic behavior revealing anomalies at 160\,K and 40\,K whose inhomogeneous character does not survive at low temperatures \cite{Poirier2014}.


In conclusion, we have demonstrated a remarkable impact of both the cation-anion coupling and van der Waals interaction on the electronic properties of QSL \EMdmit{}. The disorder identified in non-conducting cations as due to two equally probable orientations and internal rotational degrees of freedom is mapped onto the organic dimers and results in hopping dc transport and relaxor dielectric response. No sign of intra- and inter-dimer dipoles is observed, suggesting that low-lying excitations are due to the motion of charged domain walls. The ground state consists of quasi-degenerate electronic states with reduced symmetry and charge distribution in which the long range vdW force prevails over the short range Pauli repulsion. Strong correlations and vdW interactions distinguish \Betaetme3{} from \kcucn{} and \kagcn{}. Most importantly, these three organic salts display simultaneously geometrical frustration in both organic dimers, due to the triangular structures, and in the non-conducting ions. In the absence of the latter, QSL is replaced by antiferromagnetic ground state, while the relaxor peak is still observed. Hence, the disorder appears critical for the prevalence of QSL over the antiferromagnetic ordering. We expect that the present work will stimulate further studies of the microscopic understanding of this phenomenon.

\begin{acknowledgments}
We thank R.\ Kato for providing single crystals and for very useful discussions. ST and MD thank V.\ Dobrosavljevi\'{c} for stimulating discussions. This work has been supported by the Croatian Science Foundation project IP-2013-11-1011 and by the Deutsche Forschungsgemeinschaft (DFG). PL was supported by the Unity Through Knowledge Fund, Contract No.\ 22/15 and H2020 CSA Twinning Project No.\ 692194, RBI-T-WINNING. 
\end{acknowledgments}

\end{document}